# The Quest-V Separation Kernel for Mixed Criticality Systems

Ye Li, Richard West, and Eric Missimer

*Computer Science Department*
*Boston University*
*Boston, MA 02215*
Email: {liye,richwest,missimer}@cs.bu.edu

*Abstract*—Multi- and many-core processors are becoming increasingly popular in embedded systems. Many of these processors now feature hardware virtualization capabilities, such as the ARM Cortex A15, and x86 processors with Intel VT-x or AMD-V support. Hardware virtualization offers opportunities to partition physical resources, including processor cores, memory and I/O devices amongst guest virtual machines. Mixed criticality systems and services can then co-exist on the same platform in separate virtual machines. However, traditional virtual machine systems are too expensive because of the costs of trapping into hypervisors to multiplex and manage machine physical resources on behalf of separate guests. For example, hypervisors are needed to schedule separate VMs on physical processor cores. In this paper, we discuss the design of the Quest-V separation kernel, that partitions services of different criticalities in separate virtual machines, or *sandboxes*. Each sandbox encapsulates a subset of machine physical resources that it manages without requiring intervention of a hypervisor. Moreover, a hypervisor is not needed for normal operation, except to bootstrap the system and establish communication channels between sandboxes.

## I. INTRODUCTION

Embedded systems are increasingly featuring multi- and many-core processors, due in part to their power, performance and price benefits. These processors offer new opportunities for an increasingly significant class of mixed criticality systems. In mixed criticality systems, there is a combination of application and system components with different safety and timing requirements. For example, in an avionics system, the in-flight entertainment system is considered less critical than that of the flight control system. Similarly, in an automotive system, infotainment services (navigation, audio and so forth) would be considered less timing and safety critical than the vehicle management sub-systems for anti-lock brakes and traction control.

A major challenge to mixed criticality systems is the safe isolation of separate components with different levels of criticality. Isolation has traditionally been achieved by partitioning components across distributed modules, which communicate over a network such as a CAN bus. For example, Integrated Modular Avionics (IMA) [1] is used to describe a distributed real-time computer network capable of supporting applications of differing criticality levels aboard an aircraft. To implement such concepts on multi-core platform, a software architecture that enforces the safe isolation of system components is required.

This notion of component isolation, or partitioning, is the basis of software system standards such as ARINC 653 [2] and the Multiple Independent Levels of Security (MILS) [3] architecture. Current implementations of these standards into operating system kernels [4][5][6] have focused on micro-kernel and virtual machine technologies for resource partitioning and component isolation. However, due to hardware limitations and software complexity, these systems either cannot completely eliminate covert channels of communication between isolated components or add prohibitive performance overhead.

Hardware-assisted virtualization provides an opportunity to efficiently separate system components with different levels of safety, security and criticality. Back in 2006, Intel and AMD introduced their VT-x and AMD-V processors, respectively, with support for hardware virtualization. More recently, the ARM Cortex A15 was introduced with hardware virtualization capabilities, for use in portable tablet devices. Similarly, some Intel Atom chips now have VT-x capabilities for use in automobile In-Vehicle Infotainment (IVI) systems [7], and other embedded systems.

While modern hypervisor solutions such as Xen [8] and Linux-KVM [9] leverage hardware virtualization to isolate their guest systems, they are still required for CPU and I/O resource multiplexing. Expensive traps into the hypervisor occur every time a guest system needs to be scheduled, or when an I/O device transfers data to or from a guest. This situation is both unnecessary and too costly for mixed criticality systems with real-time requirements.

In this paper we present a new operating system design leveraging hardware-assisted virtualization as an extra *ring of protection*, to achieve efficient resource partitioning and performance isolation for subsystems. Our system, called Quest-V, is a separation kernel [10] design. The system avoids traps into a hypervisor (a.k.a. virtual machine monitor, or VMM) when making scheduling and I/O management decisions. Instead, all resources are partitioned at boot-time amongst system components that are capable of scheduling themselves on available processor cores. Similarly, system components are granted access to specific subsets of I/O devices and memory so that devices can be managed without involvement of a hypervisor.

In the rest of this paper, we describe how Quest-V can be configured to support a Linux front-end responsible for low criticality legacy services, while native Quest services can operate on separate hardware resources (memory, I/O and CPU cores) to ensure safe, predictable and efficient service guarantees. In this way, high criticality Quest services can co-exist with less critical Linux services on the same hardware platform.

## II. QUEST-V SEPARATION KERNEL ARCHITECTURE

Quest-V partitions a system into a series of *sandbox kernels*, with each sandbox encompassing a subset of memory, I/O and CPU resources. The current implementation works on Intel VT-x platforms but plans are underway to port Quest-V to the AMD-V and ARM architectures.

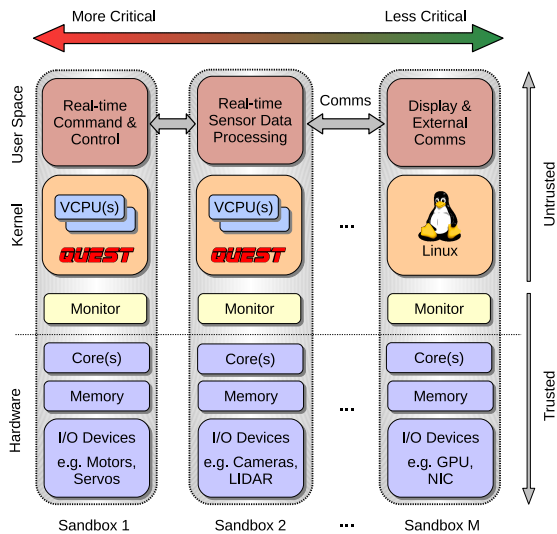

Fig. 1. Example Quest-V Architecture Overview

A high-level overview of the Quest-V architecture is shown in Figure 1. The system is partitioned into separate *sandboxes*, each responsible for a subset of machine physical memory, I/O devices and processor cores. Trusted monitor code is used to launch *guest* services, which may include their own kernels and user space programs. A monitor is responsible for managing special *extended page tables* (EPTs) that translate guest physical addresses (GPAs) to host physical addresses (HPAs), as described later in Figure 2.

We chose to have a separate monitor for each sandbox, so that it manages only one set of EPT memory mappings for a single guest environment. The amount of added overhead of doing this is small, as each monitor's code fits within $4KB$ [1]. However, the benefits are that monitors are made much simpler, since they know which sandbox they are serving rather than having to determine at runtime the guest that needs their service. Typically, guests do not need intervention of monitors, except to establish shared memory communication channels with other sandboxes, which requires updating EPTs.

**Mixed-Criticality Example** – Figure 1 shows an example of three sandboxes, where two are configured with Quest native safety-critical services for command, control and sensor data processing. These services might be appropriate for a future automotive system that assists in vehicle control. Other less critical services could be assigned to vehicle infotainment services, which are partitioned in a sandbox that has access to a local display device. A non-real-time Linux system could be used in this case, perhaps also managing a network interface (NIC) to communicate with other vehicles or the surrounding environment, via a vehicle-to-vehicle (V2V) or vehicle-to-infrastructure (V2I) communication link.

### A. Resource Partitioning

Quest-V supports configurable partitioning of CPU, memory and I/O resources amongst guests. Resource partitioning is mostly static, taking place at boot-time, with the exception of some memory allocation at runtime for dynamically created communication channels between sandboxes.

**CPU Partitioning** – In Quest-V, scheduling is performed within each sandbox. Since processor cores are statically allocated to sandboxes, there is no need for monitors to perform sandbox scheduling as is typically required with traditional hypervisors. This approach eliminates most of the monitor traps otherwise necessary for sandbox context switches. It also means there is no notion of a global scheduler to manage the allocation of processor cores amongst guests. Each sandbox's local scheduler is free to implement its own policy, simplifying resource management. This approach also distributes contention amongst separate scheduling queues, without requiring synchronization on one global queue.

---

[1]The EPTs take additional data space, but 12KB is enough for a 1GB sandbox.

**Memory Partitioning** – Quest-V relies on hardware assisted virtualization support to perform memory partitioning. Figure 2 shows how address translation works for Quest-V sandboxes using Intel's extended page tables. Each sandbox kernel uses its own internal paging structures to translate guest virtual addresses to guest physical addresses. EPT structures are then walked by the hardware to complete the translation to host physical addresses.

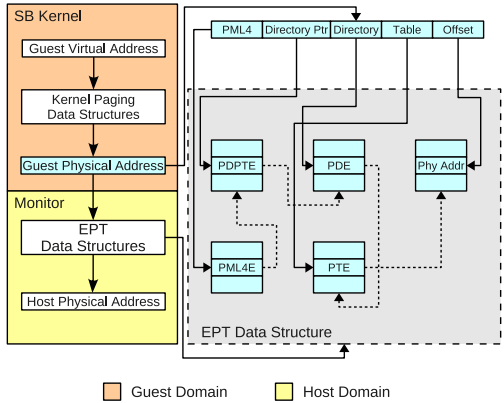

Fig. 2. Extended Page Table Mapping

On modern Intel x86 processors with EPT support, address mappings can be manipulated at 4KB page granularity. For each 4KB page we have the ability to set read, write and even execute permissions. Consequently, attempts by one sandbox to access illegitimate memory regions of another will incur an EPT violation, causing a trap to the local monitor. The EPT data structures are, themselves, restricted to access by the monitors, thereby preventing tampering by sandbox kernels.

EPT mappings are cached by hardware TLBs, expediting the cost of address translation. Only on returning to a guest after trapping into a monitor are these TLBs flushed. Consequently, by avoiding exits into monitor code, each sandbox operates with similar performance to that of systems with conventional page-based virtual address spaces [11].

**I/O Partitioning** – In Quest-V, device management is performed within each sandbox directly. Device interrupts are delivered to a sandbox kernel without monitor intervention. This differs from the "split driver" model of systems such as Xen, which have a special domain to handle interrupts before they are directed into a guest. Allowing sandboxes to have direct access to I/O devices greatly reduces the overhead of monitor traps to handle interrupts.

To partition I/O devices, Quest-V first has to restrict access to device specific hardware registers. Device registers are usually either memory mapped or accessed through a special I/O address space (e.g. I/O ports). For the x86, both approaches are used. For memory mapped registers, EPTs are used to prevent their accesses from unauthorized sandboxes. For port-addressed registers, special hardware support is necessary. On Intel processors with VT-x, all variants of `in` and `out` instructions can be configured to cause a monitor trap if access to a certain port address is attempted. As a result, a hardware provided I/O bitmap can be used to partition the whole I/O address space amongst different sandboxes. Unauthorized access to a certain register can thus be ignored or trigger a fault recovery event.

Any sandbox attempting access to a PCI device must use memory-mapped or port-based registers identified in a special PCI *configuration space* [12]. Quest-V intercepts access to this configuration space, which is accessed via both an address and data I/O port. A trap to the local sandbox monitor occurs when there is a PCI data port access. The monitor then determines which device's configuration space is to be accessed by the trapped instruction. A device *blacklist* for each sandbox containing the *Device ID* and *Vendor ID* of restricted PCI devices is used by the monitor to control actual device access.

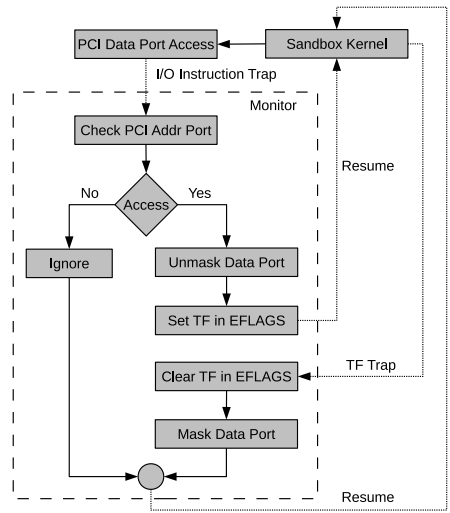

Fig. 3. PCI Configuration Space Protection

A simplified control flow of the handling of PCI configuration space protection in a Quest-V monitor is given in Figure 3. Notice that simply allowing access to a PCI data port is not sufficient because we only want to allow the single I/O instruction that caused the monitor trap, and which passed the monitor check, to be correctly executed. Once this is done, the monitor should immediately restrict access to the PCI data port again. This behavior is achieved by setting the *trap flag* (TF) bit in the sandbox kernel system flags to cause a single step debug exception after it executes the next instruction. By configuring the processor to generate a monitor trap on debug exception, the system can immediately return to the monitor after executing the I/O instruction. After this, the monitor is able to mask the PCI data port again for the sandbox kernel, thereby mediating future device access.

In addition to direct access to device registers, interrupts from I/O devices also need to be partitioned amongst sandboxes. In modern multi-core platforms, an external interrupt controller is almost always present to allow configuration of interrupt delivery behaviors. On modern Intel x86 processors, this is done through an I/O Advanced Programmable Interrupt Controller (IOAPIC). Each IOAPIC has an I/O *redirection table* that can be programmed to deliver device interrupts to all, or a subset of, sandboxes. Each entry in the I/O redirection table corresponds to a certain interrupt request from an I/O device on the PCI bus.

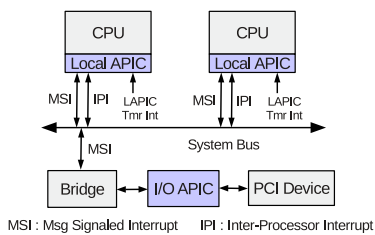

Fig. 4. APIC Configuration

Figure 4 shows the hardware APIC configuration. Quest-V uses EPT entries to restrict access to memory regions, used to access IOAPIC registers. This way, the IOAPIC interrupt redirection table is prevented from alteration by an unauthorized sandbox. Attempts by a sandbox to access the IOAPIC space cause a trap to the local monitor as a result of an EPT violation. The monitor then checks to see if the sandbox has authorization to update the table before allowing any changes to be made. Consequently, device interrupts are safely partitioned amongst sandboxes.

### B. Linux Sandbox Support

In addition to native Quest kernels, Quest-V is also designed to support other third party sandbox kernels such as Linux and AUTOSAR OS. Currently, we have successfully ported a Puppy Linux distribution with Linux 3.8 kernel to serve as our system front-end, providing a window manager and graphical user interface. A Quest kernel is first started in the Linux sandbox to bootstrap our paravirtualized Linux kernel. Because Quest-V exposes maximum possible privilege of hardware access to sandbox kernels, the actual changes made to the original Linux 3.8 kernel are well under 100 lines. These changes are mainly focused on limiting Linux's view of available memory and handling I/O device DMA offsets caused by memory virtualization.

The VGA frame buffer and GPU hardware are always assigned to the bootstrapped Linux sandbox. All the other sandboxes will have their default terminal I/O tunneled through shared memory channels to virtual terminals in the Linux front-end. We developed libraries, user space applications and a kernel module to support this redirection in Linux. A screen shot of Quest-V after booting the Linux front-end sandbox is shown in Figure 5. Here, we show two virtual terminals connected to two different native Quest sandboxes similar to the configuration shown in Figure 1. In this particular example, we allocated 512MB of memory to the Linux sandbox and 256MB of memory to each native Quest sandbox. The network interface card has been assigned to Quest sandbox 1, while the serial device has been assigned to Quest sandbox 2. The Linux sandbox is granted ownership of the USB host controller in addition to the GPU and VGA frame buffer. Observe that although the machine has four processor core, the Linux kernel detects only one core.

### C. VCPU Scheduling

Sandboxes running Quest kernels use a form of *virtual CPU* (VCPU) scheduling for conventional tasks and interrupt handlers [13]. The concept of a VCPU is similar to that in traditional virtual machines [8], where a hypervisor provides the illusion of multiple *physical CPUs* (PCPUs) [2] represented as VCPUs to each of the guests. VCPUs exist as kernel abstractions to simplify the management of resource budgets for potentially many software threads. We use a hierarchical approach in which VCPUs are scheduled on PCPUs and threads are scheduled on VCPUs.

A VCPU acts as a resource container [14] for scheduling and accounting decisions on behalf of software threads. It serves no other purpose to virtualize the underlying physical CPUs, since our sandbox kernels and their applications execute directly on the hardware. In particular, a VCPU does not need to act as a container for cached instruction blocks that have been generated to emulate the effects of guest code, as in some trap-and-emulate virtualized systems.

---
[2]We define a PCPU to be either a conventional CPU, a processing core, or a hardware thread.

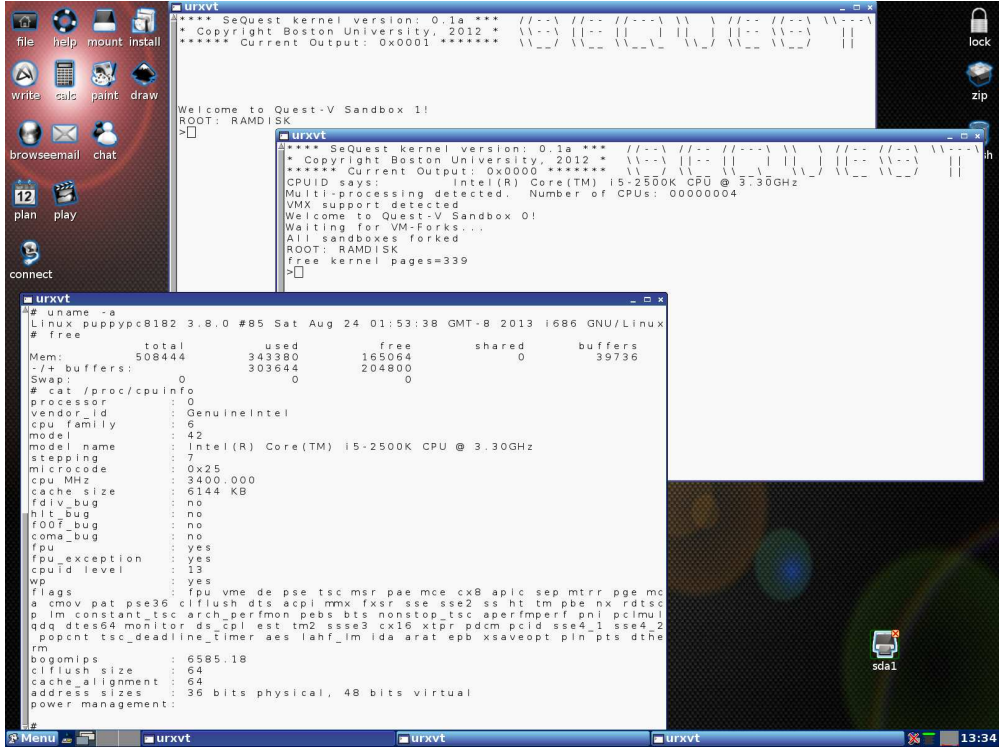

Fig. 5. Quest-V with Linux Front-End

In common with bandwidth preserving servers [15], [16], [17], each VCPU, $V$, has a maximum compute time budget, $C_{max}$, available in a time period, $V_T$. $V$ is constrained to use no more than the fraction $V_U = \frac{C_{max}}{V_T}$ of a physical processor (PCPU) in any window of real-time, $V_T$, while running at its normal (foreground) priority. To avoid situations where PCPUs are idle when there are threads awaiting service, a VCPU that has expired its budget may operate at a lower (background) priority. All background priorities are set below those of foreground priorities to ensure VCPUs with expired budgets do not adversely affect those with available budgets.

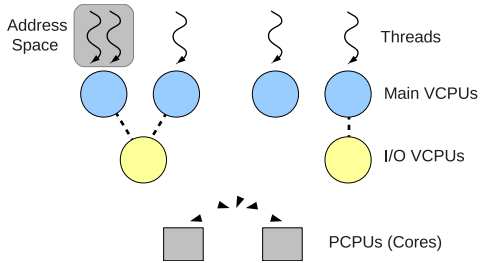

Fig. 6. VCPU Scheduling Hierarchy

Quest kernels define two classes of VCPUs as shown in Figure 6: (1) *Main VCPUs* are used to schedule and track the PCPU usage of conventional software threads, while (2) *I/O VCPUs* are used to account for, and schedule the execution of, interrupt handlers for I/O devices. This distinction allows for interrupts from I/O devices to be scheduled as threads, which may be deferred execution when threads associated with higher priority VCPUs having available budgets are runnable. I/O VCPUs can be specified for certain devices, or for certain tasks that issue I/O requests, thereby allowing interrupts to be handled at different priorities and with different CPU shares than conventional tasks associated with Main VCPUs.

By default, VCPUs act like Sporadic Servers [18], [19]. Local APIC timers are programmed to replenish VCPU budgets as they are consumed during thread execution. Sporadic Servers enable a system to be treated as a collection of equivalent periodic tasks scheduled by a rate-monotonic scheduler (RMS) [20]. This is significant, given I/O events can occur at arbitrary (aperiodic) times, potentially triggering the wakeup of blocked tasks (again, at arbitrary times) having higher priority than those currently running. RMS analysis can be applied, to ensure each VCPU is guaranteed its share of CPU time, $V_U$, in finite windows of real-time.

## III. EXISTING SOLUTIONS AND RELATED WORK

Xen [8], Linux-KVM [9], XtratuM [6] and the Wind River Hypervisor [21] all use virtualization technologies to logically isolate and multiplex guest virtual machines on a shared set of physical resources. LynxSecure [5] is another similar approach targeted at safety-critical real-time systems.

PikeOS [4] is a separation micro-kernel [22] that supports multiple guest VMs, and targets safety-critical domains such as Integrated Modular Avionics. The micro-kernel supports a virtualization layer that is required to manage the spatial and temporal partitioning of resources amongst guests.

In contrast to the above systems, Quest-V statically partitions machines resources into separate sandboxes. Services of different criticalities can be mapped into separate sandboxes. Each sandbox manages its own resources independently of an underlying hypervisor. Quest-V also avoids the need for a split-driver model involving a special domain (e.g., `Dom0` in Xen) to handle device interrupts. Interrupts are delivered directly to the sandbox associated with the corresponding device, using I/O passthrough.

## IV. CONCLUSIONS AND FUTURE WORK

This paper introduces Quest-V, which is an open-source separation kernel built from the ground up. It uses hardware virtualization in a first-class manner to separate system components of different criticalities. It avoids traditional costs associated with hypervisor systems, by statically allocating partitioned subsets of machine resources to guest sandboxes, which perform their own scheduling, memory and I/O management. Hardware virtualization simply provides an extra logical ring of protection. Sandboxes can communicate via shared memory channels that are mapped to immutable entries in extended page tables (EPTs). Only trusted monitors are capable of changing the entries in these EPTs, preventing access to arbitrary memory regions in remote sandboxes.

Quest-V requires system monitors to be trusted. Although these occupy a small memory footprint and are not involved in normal system operation, the system can be compromised if the monitors are corrupted. Future work will investigate the use of hardware features such as Intel's trusted execution technology (TXT) to enforce safety of the monitors. Additionally, online fault detection and recovery strategies will be considered.

Please see `www.questos.org` for more details.